Photometry of Kuiper Belt Object (486958) Arrokoth from New Horizons LORRI

Jason D. Hofgartner*[1], Bonnie J. Buratti[1], Susan D. Benecchi[2], Ross A. Beyer[3,4], Andrew Cheng[5], James T. Keane[6], Tod R. Lauer[7], Catherine B. Olkin[8], Joel W. Parker[8], Kelsi N. Singer[8], John R. Spencer[8], S. Alan Stern[8], Anne J. Verbiscer[9], Harold A. Weaver[5], New Horizons Geology and Geophysics Team, and New Horizons LORRI Team

(*) Jason.D.Hofgartner@jpl.nasa.gov
(1) Jet Propulsion Laboratory, California Institute of Technology, California, USA, (2) Planetary Science Institute, Arizona, USA, (3) Sagan Center at SETI Institute, California, USA, (4) NASA Ames Research Center, California, USA, (5) Johns Hopkins University Applied Physics Laboratory, Maryland, USA, (6) California Institute of Technology, California, USA, (7) NSF's National Optical Infrared Astronomy Research Laboratory, Arizona, USA, (8) Southwest Research Institute, Colorado, USA, (9) University of Virginia, Virginia, USA

**Abstract**

On January 1st 2019, the New Horizons spacecraft flew by the classical Kuiper belt object (486958) Arrokoth (provisionally designated 2014 MU69), possibly the most primitive object ever explored by a spacecraft. The I/F of Arrokoth is analyzed and fit with a photometric function that is a linear combination of the Lommel-Seeliger (lunar) and Lambert photometric functions. Arrokoth has a geometric albedo of $p_v = 0.21^{+0.05}_{-0.04}$ at a wavelength of 550 nm and ≈ 0.24 at 610 nm. Arrokoth's geometric albedo is greater than the median but consistent with a distribution of cold classical Kuiper belt objects whose geometric albedos were determined by fitting a thermal model to radiometric observations. Thus, Arrokoth's geometric albedo adds to the orbital and spectral evidence that it is a cold classical Kuiper belt object. Maps of the normal reflectance and hemispherical albedo of Arrokoth are presented. The normal reflectance of Arrokoth's surface varies with location, ranging from ≈ 0.10 – 0.40 at 610 nm with an approximately Gaussian distribution. Both Arrokoth's extrema dark and extrema bright surfaces are correlated to topographic depressions. Arrokoth has a bilobate shape and the two lobes have similar normal reflectance distributions: both are approximately Gaussian, peak at ≈ 0.25 at 610 nm, and range from ≈ 0.10 – 0.40, which is consistent with co-formation and co-evolution of the two lobes. The hemispherical albedo of Arrokoth varies substantially with both incidence angle and location, the average hemispherical albedo at 610 nm is 0.063 ± 0.015. The Bond albedo of Arrokoth at 610 nm is 0.062 ± 0.015.

# 1. Introduction

On January $1_{st}$ 2019, the New Horizons spacecraft approached within 3500 km of (486958) Arrokoth (provisionally designated 2014 MU69 and informally named Ultima Thule; Stern et al., 2019), hereafter Arrokoth, possibly the most primitive object ever explored by a spacecraft. Arrokoth is a Kuiper belt object (KBO) with a semimajor axis of ≈ 44.2 au and eccentricity of ≈ 0.04 (Porter et al., 2018) and thus the solar energy incident on its surface is weak. The cumulative impacts on its surface over Solar System history were predicted to be relatively low in both abundance and speed, such that the primordial surface may not be saturated by craters (Greenstreet et al., 2019), consistent with New Horizons' observation of a low crater density (Stern et al., 2019; Spencer et al., accepted; Singer et al., 2019). The volume-equivalent spherical diameter of Arrokoth is ≈ 18 km, so internal heating from accretion and radionuclides is expected to be weak (Stern et al., 2019; Spencer et al., accepted). Thus, the surface has most likely experienced relatively little modification from solar energy, impacts, and internal energy and is likely more pristine than other objects explored by spacecraft. However, the primordial surface of Arrokoth has experienced space weathering (Pieters and Noble, 2016) since the epoch of its formation.

The Kuiper belt has several dynamical sub-populations (Petit et al., 2011) and based on its orbit, Arrokoth is likely a cold classical Kuiper belt object (CCKBO) and a member of the kernel (Porter et al., 2018), the sub-population that has experienced the least dynamical perturbations and likely formed at its current location. Arrokoth is therefore also dynamically primitive.

Ground-based observations suggested (Buie et al., submitted) and New Horizons confirmed Arrokoth to be a bilobate object (Stern et al., 2019). The connection zone between the two lobes is brighter than its surroundings and is referred to as the neck (Stern et al., 2019). Arrokoth has an obliquity of ≈ $99_o$ and as a result, most of the northern hemisphere of Arrokoth was not illuminated and imaged by New Horizons (Spencer et al., accepted). The smaller lobe is dominated by an ≈ 7 km diameter impact crater that is almost an order of magnitude larger than the next largest crater observed on Arrokoth; it is informally referred to as Maryland (Spencer et al., accepted; Singer et al., 2019). Arrokoth has a nearly uniform red color at visible wavelengths that is consistent with other CCKBOs (Benecchi et al., 2019b; Grundy et al., accepted). The surface is composed of organic macromolecules (similar to tholins produced in terrestrial laboratories), amorphous carbon, and methanol; surprisingly, water-ice absorption bands are weak or not present in the reflectance spectrum (Grundy et al., accepted). No satellites or rings orbiting Arrokoth were discovered (Stern et al., 2019; Spencer et al., accepted).

We determine the normal reflectance, geometric albedo, hemispherical albedo, and Bond albedo of Arrokoth's surface. Most variations of the observed brightness of a surface are not intrinsic, but rather due to variation of the observation geometry, which is defined by the incident, emission, and solar phase angles (photometric angles; e.g., Hofgartner et al., 2018). Normal reflectance is the I/F (where I is the scattered intensity from the surface and $\pi F$ is the solar flux at the distance of the scattering surface; also called the radiance factor (e.g., Hapke, 2012)) when these three photometric angles are zero degrees; it is a measure of the intrinsic brightness of a surface. Geometric albedo is the disk-integrated I/F (note that since Arrokoth is not spherical, its projected shape is not a disk; we use the term *disk-integrated* in the generalized sense of integration over the projected area) at a solar phase angle of zero degrees (opposition). Geometric albedo is a disk-integrated quantity whereas normal reflectance is an analogous spatially-resolved quantity. Hemispherical albedo is the ratio of the total power scattered by a surface to the incident power and is crucial for understanding the thermal evolution of the surface. Bond albedo is the analogous disk-integrated albedo; it is the ratio of the total power scattered by a planetary body to the incident power. Note that for both the hemispherical albedo and Bond albedo, total power refers to the total in an angular sense (i.e., integration over all emission angles); bolometric hemispherical albedo and bolometric Bond albedo are the ratios for the total power in both angular and spectral senses (i.e., integration over all emission angles and all wavelengths). All of these albedos provide clues about the properties of a surface and its geology. We present maps of the normal reflectance and hemispherical albedo of Arrokoth. We analyze the normal reflectance distribution over Arrokoth's surface and compare our results for Arrokoth to other KBOs as well as Centaurs, comets, and irregular satellites, all or some of which may be former KBOs.

In section 2, the photometric function we use to correct for brightness variations due to changes of the photometric angles is described. A map of the normal reflectance of Arrokoth is presented in section 3 and its geometric albedo is determined. Section 4 discusses the hemispherical and Bond albedos of Arrokoth. The photometry of Arrokoth is compared to cognate Solar System objects in section 5. The conclusions are provided in the final section.

## 2. Photometric Function

Various photometric functions describe the brightness of planetary surfaces as a function of the photometric angles (incidence, emission, and solar phase; e.g., Hapke, 2012). For New Horizons' observations of Arrokoth, we use a linear combination of Lommel-Seeliger (also known as lunar) and Lambert photometric functions and refer to the combined function as the lunar-Lambert function. The Lommel-Seeliger function is an analytic solution to the equations of radiative transfer when multiple scattering is ignored. The Lambert function describes perfectly diffuse scattering and is a good approximation when the surface reflectance is dominated by multiple scattering. The empirical lunar-Lambert function has been used to study several planetary surfaces, particularly in the outer Solar System (e.g., Buratti and Veverka, 1983; Buratti et al., 2017), and is appropriate for a limited dataset because it has few parameters.

The equation we use for the lunar-Lambert photometric function is:

$$\frac{I}{F} = A \frac{f(\alpha) \cos i}{\cos i + \cos e} + (1 - A) \cos i \qquad (1)$$

where $I$ is the scattered intensity from the surface of Arrokoth, $F$ is the solar flux at Arrokoth divided by pi, the first term on the right side of the equation is the Lommel-Seeliger (lunar, single scattering) photometric function and the second term is the Lambert (diffuse, multiple scattering) function. The photometric angles are $i, e,$ and $\alpha$, where $i$ is incidence angle (angle at the surface between the direction to the Sun and the surface normal), $e$ is emission angle (angle at the surface between the direction to the camera and the surface normal) and $\alpha$ is solar phase angle (Sun-surface-camera angle). $A$, is an empirical parameter that depends on the relative contributions of the lunar and Lambert functions and the magnitude of the normal reflectance and $f(\alpha)$ is the surface phase function, which depends on physical properties such as surface roughness and compaction. We note that $A$ is not the partition between the lunar and Lambert functions, because it also incorporates the total magnitude of the surface brightness (for example, $A = 0.5$ does not imply that the lunar and Lambertian terms contribute equally to the I/F; McEwen, 1986). The lunar-Lambert function can be expressed in slightly different forms than equation 1 (e.g., McEwen, 1986). For example, $A$ could be allowed to depend on solar phase angle; however, in that case the second term is no longer the Lambertian function for diffuse scattering that is well known in planetary photometry, so we treat $A$ as a constant. Equation 1 is valid for $0º \leq i \leq 90º$ and $0º \leq e \leq 90º$, otherwise $I/F = 0$.

To determine the photometric angles at each location on the surface, the surface shape must be known. The shape can often be approximated as a sphere, but that approximation is inadequate for Arrokoth's complex shape; two-spheres is also insufficient. To determine the photometric angles at each location on the surface of Arrokoth, we used a merged shape model (Spencer et al., accepted) that combines a global, low-resolution shape model (Porter et al., 2019), and a stereo topographic model of the ventral surface of Arrokoth (Beyer et al., 2019). This merged model was created by fitting the stereo model to the shape model, and then eroding both models back from their intersection, joining them with polygons, and then locally smoothing the model at the join (Beyer et al., 2019).

The New Horizons spacecraft approached Arrokoth from a solar phase angle asymptote of ≈ 11.8º. New Horizons' narrow angle camera, the Long Range Reconnaissance Imager (LORRI; Cheng et al., 2008), acquired many images at this solar phase angle with increasingly better resolution as the spacecraft neared Arrokoth. In this analysis, we use the best resolution LORRI images at $\alpha \approx$ 11.8º and all subsequent LORRI images that detected Arrokoth, except a crescent image at $\alpha \approx$ 150º because the available merged shape model did not adequately fit the observed crescent to reliably predict the incidence and emission angles. This corresponds to five image sequences referred to as CA01, CA02, CA04, CA05, and CA06; the solar phase angle and LORRI pixel scale for these image sequences are provided in Table 1 (additional observational details are provided in the supplementary material of Spencer et al., accepted). Each of these image sequences corresponds to a series of images acquired in rapid succession. The images in each series were pipeline processed as described in Weaver et al., submitted, then deconvolved and stacked using the techniques described in Weaver et al., 2016, then the mode of the sky background was subtracted from the whole image. We analyze the resulting stacked, background-subtracted images.

Table 1: Stacked New Horizons LORRI images that are analyzed, their solar phase angles and pixel scales, and best-fit surface phase function ($f(\alpha)$) for the lunar-Lambert photometric function.

| Stacked Images | Solar Phase Angle (degrees) | Pixel Scale (m/pixel) | Best-fit surface phase function for lunar-Lambert photometric function |
|---|---|---|---|
| CA01 | 11.8 | 300 | $f(11.8º) = 0.109$ |
| CA02 | 12.0 | 212 | $f(12.0º) = 0.111$ |
| CA04 | 13.0 | 137 | $f(13.0º) = 0.105$ |
| CA05 | 15.7 | 83 | $f(15.7º) = 0.098$ |
| CA06 | 32.6 | 33 | $f(32.6º) = 0.057$ |

Figure 1A shows the measured I/F of Arrokoth in all five images as a function of the photometric angles. Each image corresponds to a plane with an approximately constant solar phase angle. Figure 1B shows the measurements for only the CA06 image. Systematic variations with each photometric angle are observed in both panels. The parameters of the best-fit lunar-Lambert function to all of the data in figure 1A are $A = 0.970$, $f(11.8º) = 0.109$, $f(12.0º) = 0.111$, $f(13.0º) = 0.105$, $f(15.7º) = 0.098$, $f(32.6º) = 0.057$. This global fit to the full dataset implicitly assumes that the photometric behavior is the same for all locations on the surface. Figures 1C and 1D show the residuals after subtracting the best-fit lunar-Lambert function. The trends with the photometric angles are no longer present, indicating that the best-fit photometric function adequately describes the photometric behavior of Arrokoth. Variations within the residual values of each image correspond primarily to intrinsic brightness differences between different locations but also errors of the photometric angles (from an imperfect shape model (Porter et al., 2019; Beyer et al., 2019)) and noise in the measured I/F.

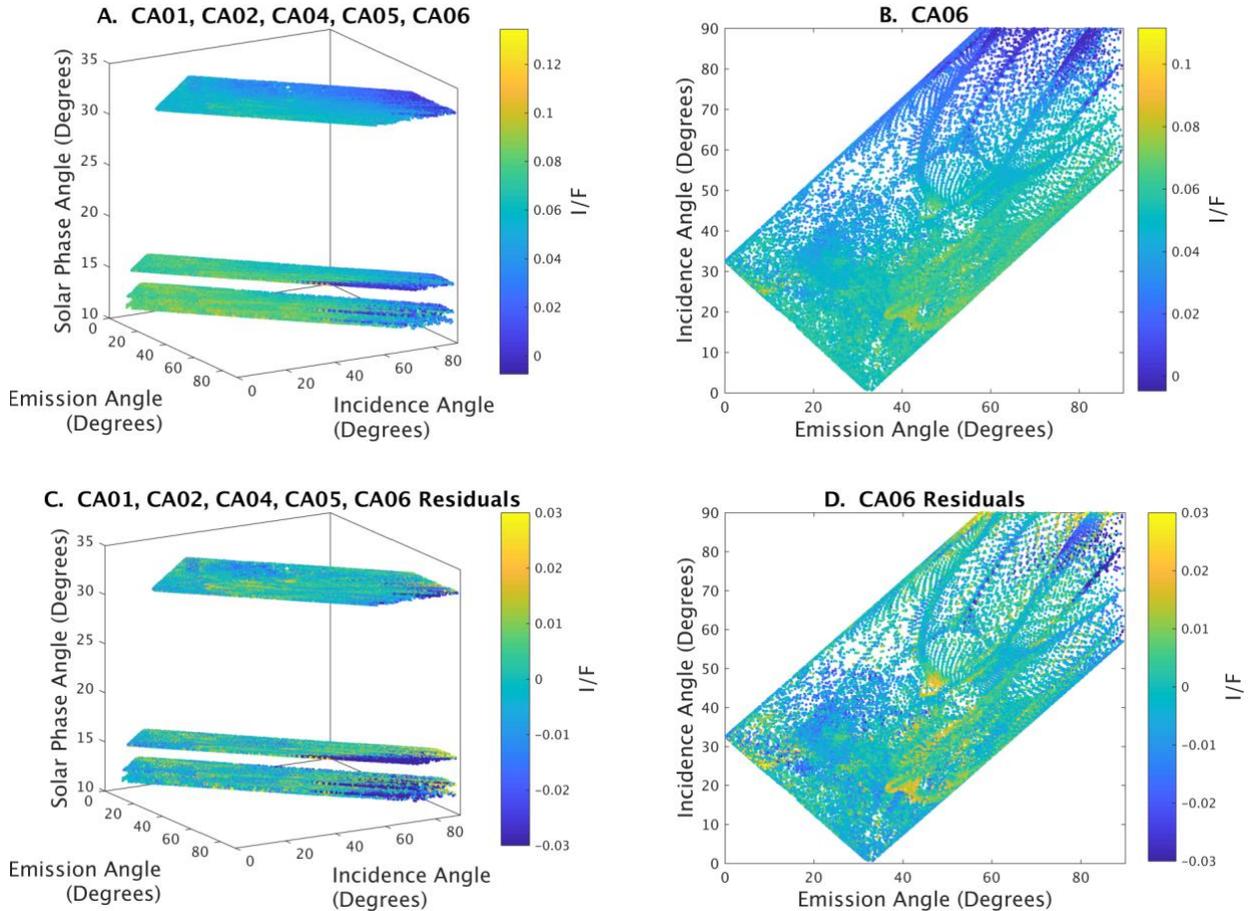

Figure 1: A. I/F (radiance factor) of (486958) Arrokoth's surface as a function of the photometric angles. Each plane with approximately constant solar phase angle corresponds to one of the five images. Each I/F measurement (dot) corresponds to a pixel that includes Arrokoth's surface. Systematic trends of the I/F with each photometric angle are observed. B. I/F as a function of incidence and emission angles for the CA06 image. The heterogenous sampling of the emission angle and incidence angle phase space is due to Arrokoth's irregular shape and topography. Approximately 0.3% of the I/F values in both A. and B. are < 0; this is a result of noise and pipeline processing of the images. C. Residuals after the best-fit lunar-Lambert function is subtracted from the data in panel A. The systematic trends with the photometric angles are no longer apparent, indicating that the photometric function accounts for the effect of the photometric angles on surface

brightness. The remaining variability in the residuals is primarily due to intrinsic brightness differences between different locations. D. Residuals after the best-fit photometric function is subtracted from the data in panel B.

## 3. Normal Reflectance and Geometric Albedo of (486958) Arrokoth

Normal reflectance is the I/F when the incidence, emission, and solar phase angles are zero degrees and is a measure of the intrinsic brightness of a surface. The normal reflectance of each location on Arrokoth's surface can be determined using the best-fit lunar-Lambert photometric function:

$$r_{n;j} = \left(\frac{I}{F}(\alpha, i, e)\right)_j \left(\frac{A\frac{f(0)}{2} + (1 - A)}{A\frac{f(\alpha)\cos i}{\cos i + \cos e} + (1 - A)\cos i}\right) \quad (2)$$

where $r_{n;j}$ is the normal reflectance of (pixel) location $j$. The above equation assumes that the ratio of the normal reflectance to the observed I/F at photometric angles $\alpha, i, e$ is the same for all locations on the surface, consistent with the earlier assumption that a single photometric function is applicable at all locations. $(I(\alpha, i, e)/F)_j$ is the measured I/F at (pixel) location $j$ in the image, $A$ and $f(\alpha)$ are determined by the best-fit function. The parameter $f(0)$, however, is not constrained by the images since New Horizons' did not image Arrokoth at $\alpha < 11.8°$. This is a common limitation of spacecraft data, especially for flyby missions, and one solution is to use Earth-based observations at $\alpha \approx 0°$.

The opposition ($\alpha = 0°$) magnitude from Earth-based observations can be expressed as a disk-integrated I/F; the disk-integrated I/F at opposition is called the geometric albedo. The equation for determining the geometric albedo is:

$$p = 10^{0.4(m_{sun}-H)} \frac{\pi r_{au}^2}{a} \quad (3)$$

where $p$ is the geometric albedo, $m_{sun}$ is the apparent magnitude of the Sun, $H$ is the absolute magnitude of Arrokoth, $r_{au}$ is the distance of one astronomical unit, and $a$ is the projected area of Arrokoth. The observed geometric albedo can then be equated to the expected geometric albedo for the lunar-Lambert photometric function to determine $f(0)$. For a sphere, the expected geometric albedo is an analytic expression (e.g., Buratti and Veverka, 1983). For a complex shape such as that of Arrokoth, however, the geometric albedo for a lunar-Lambert photometric function is not a simple analytic expression. Using the merged shape model for Arrokoth, the best-fit lunar-Lambert function, and assuming a value for $f(0)$, we simulate the geometric albedo. We repeat the simulation for a range of $f(0)$ and fit the predicted geometric albedo to the observed geometric albedo from Earth-based observations, effectively fitting for $f(0)$.

The absolute magnitude of Arrokoth was determined using the Hubble Space Telescope F350LP filter (Benecchi et al., 2019a) and we calculate that the weighted mean of all the reported measurements is $H_{F350LP} = 10.47$. Using the observed color of Arrokoth (Benecchi et al., 2019b) this corresponds to a Johnson V-band absolute magnitude in the Vega-magnitude system of $H_V = 10.40$. The apparent solar magnitude in this system is -26.76 (Willmer, 2018). The rotationally averaged projected area, to the Sun, in the middle of 2016 (approximate midpoint of the Earth-based observations to measure Arrokoth's absolute magnitude (Benecchi et al., 2019a)) of Arrokoth was $4.57 \times 10^8$ m2. Based on these values and the above equation, we calculate that the geometric albedo of Arrokoth in the V-band (pivot wavelength of 551.1 nm) is ≈ 0.21. The uncertainty of the geometric albedo is dominated by the uncertainty of the absolute magnitude of Arrokoth. To estimate the uncertainty, we use the standard deviation of the Hubble Space Telescope measurements, which gives $p_v = 0.21^{+0.05}_{-0.04}$. Note that the geometric albedo does not depend linearly on the absolute magnitude ($H$), so the uncertainty does not simply double and triple for two standard deviations (2-sigma) and three standard deviations (3-sigma). Also, the statistical uncertainty of the weighted mean of the absolute magnitude is < 1/20th the standard deviation of the measurements; however, we consider the larger uncertainty stated above to be a better estimate due to possible systematic errors.

This V-band geometric albedo of $0.21^{+0.05}_{-0.04}$ is greater than that reported for Arrokoth in Stern et al., 2019 (0.165 ± 0.01), but consistent within the uncertainties, because that work used an absolute magnitude of $H_{F350LP} = 10.86$; 10.47 is the most-current measurement (Benecchi et al., 2019a). Note that the absolute values of the geometric albedo and normal reflectances that we report in this work depend strongly on the V-band absolute magnitude, which was not measured by New Horizons. If the measured value of the absolute magnitude changes, the values of the geometric albedo and normal reflectances will change accordingly. We do not expect the visible absolute magnitude to differ substantially from the color-adjusted, weighted mean we calculated above, from the measurements in Benecchi et al., 2019a; 2019b, but are explicitly acknowledging this sensitivity. The relative variations of the normal reflectance of Arrokoth's surface in the normal reflectance map, however, are not affected by the absolute magnitude.

Using the measured color slope of Arrokoth (Grundy et al., accepted), we determined the geometric albedo at the LORRI pivot wavelength (607.6 nm; Cheng et al., 2008) to be $\approx 0.24$. We simulated the geometric albedo of Arrokoth in the middle of 2016 (using its geometry relative to the Sun and a hypothetical observer at $\alpha = 0_o$). We repeated the simulation for 12 different subsolar longitudes to account for rotational variability and then averaged the results. The Arrokoth geometric albedo of 0.24 is best-fit by the simulations with $f(0) = 0.456$.

Figure 2 shows the normal reflectance map of Arrokoth. Since Arrokoth is a complex shape and most of the northern hemisphere was not imaged (a result of Arrokoth's high obliquity and New Horizons' approach from the inner Solar System), the map is not projected but displayed with the same geometry as the CA06 image. The map for the CA01 image geometry is also included to show some regions of the surface that were imaged by New Horizons but are not apparent in the CA06 geometry, albeit at lower spatial resolution (Table 1). The normal reflectance varies across the surface of Arrokoth. The darkest region is in the Maryland crater on the smaller lobe and has a normal reflectance of $\approx 0.10$. A depression on the larger lobe, near the neck, on the left side of the CA06 image in figure 2 (this area is informally referred to as Louisiana), has similarly low normal reflectance. The brightest normal reflectance of Arrokoth is $\approx 0.45$ and occurs at the neck, two bright spots in the Maryland crater, and two small spots on the larger lobe near the neck and on the right side of the CA06 image in figure 2.

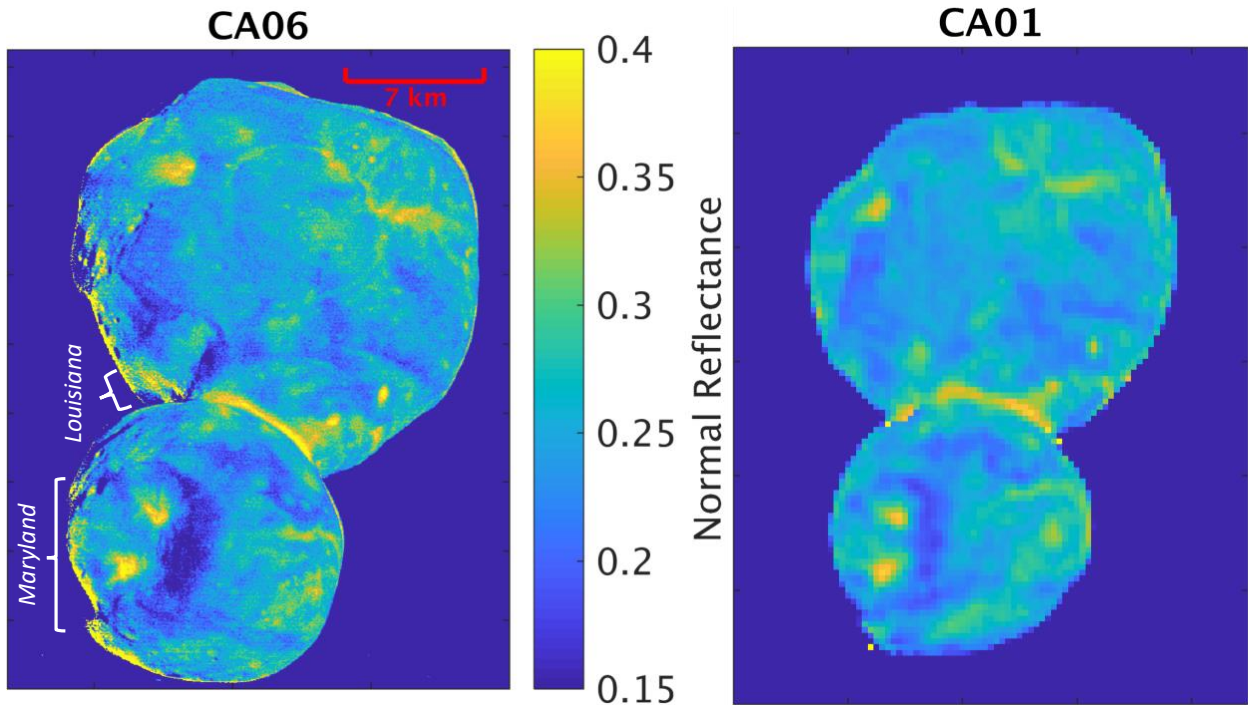

Figure 2: Normal reflectance map of (486958) Arrokoth at New Horizons LORRI pivot wavelength of 607.6 nm. The left panel is the CA06 image, the highest-resolution image from the New Horizons flyby, corrected to normal reflectance using the lunar-Lambert photometric function. The normal reflectance map for the CA01 image is also included in the right panel for the sake of completeness, since it includes some areas not visible in the CA06 image. The color scale bar applies to both images. The regions that are informally referred to as Maryland and Louisiana are indicated in the figure. Anomalously bright edges at the outline of Arrokoth are artifacts.

The brightest regions of Arrokoth are correlated to depressions, suggesting that the topography influences the albedo and/or vice versa. A possible explanation is that volatiles have accumulated in the depressions, but we consider this explanation unlikely since the normal reflectances are not as high as that of known volatile-rich surfaces in the Kuiper belt (e.g., Buratti et al., 2017; Hofgartner et al., 2019) and the surface temperatures of these regions should differ from their surroundings by only a few Kelvin (Grundy et al., accepted). Another possible explanation is that topographic shielding of energetic radiation that would chemically modify the surface toward dark-red material (similar to tholins produced in terrestrial laboratories) inhibits this processing in depressions. If true, it is surprising that these surfaces are not as processed as their surroundings, given that they likely also date from the epoch of formation. The hypothesis that we favor is accumulation of fine grains in the depressions, possibly from transport across the surface. A more detailed investigation of these hypotheses is warranted but left to future work. The correlation between high normal reflectance and low topography also suggests that the neck connecting the two lobes of Arrokoth may not be anomalous but only the largest example of Arrokoth's bright depressions; the neck may not be bright due to a process that operated (or operates) solely on the neck. Intriguingly, the darkest observed regions on both lobes are also in depressions, the largest observed depression on each lobe.

Figure 3A shows the normal reflectance distribution of Arrokoth's surface. The distribution is unimodal with a mode at $\approx 0.25$ (recall that this is at the LORRI pivot wavelength of 607.6 nm) and approximately symmetric and Gaussian. The approximately Gaussian distribution of Arrokoth's normal reflectance indicates that the geometric albedo is unlikely to be substantially skewed because of any particular terrain type, such as the bright neck connecting the two lobes. The normal reflectance distributions of the individual lobes are also included in figure 3A. The distributions of the lobes are similar to that of the whole of Arrokoth, they are approximately symmetric and Gaussian, and peak at $\approx 0.25$. The best-fit normalized Gaussian distributions are shown in figure 3B; only normal reflectances from 0.15-0.35 were included in the fits to avoid broadening of the distributions by values at the wings (outside of this range). The best-fit mean and standard deviation for Arrokoth's surface are 0.249 and 0.036, 0.251 and 0.034 for the larger lobe, and 0.244 and 0.040 for the smaller lobe. The distribution of the smaller lobe is broader and skewed to lower normal reflectance. These differences, however, are not apparent when the large Maryland crater is not included in the histogram for the smaller lobe, as shown in figure 3C. Thus, the two lobes have very similar distributions, but the smaller lobe's distribution is slightly different due to its large crater.

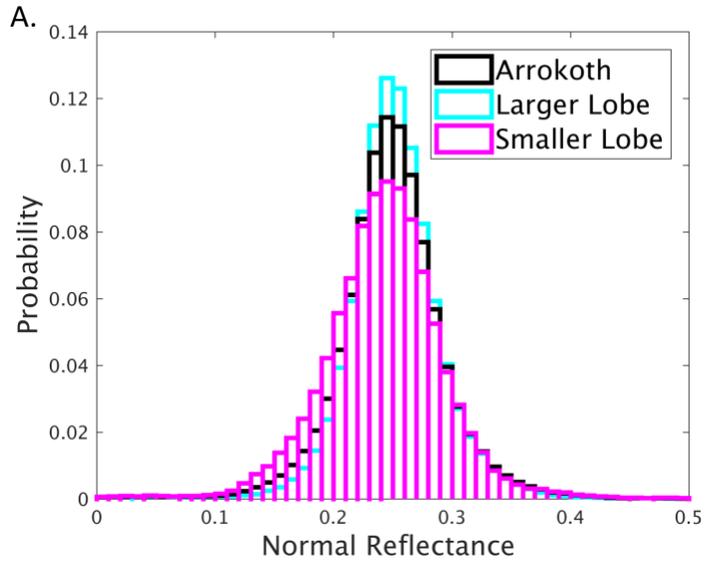
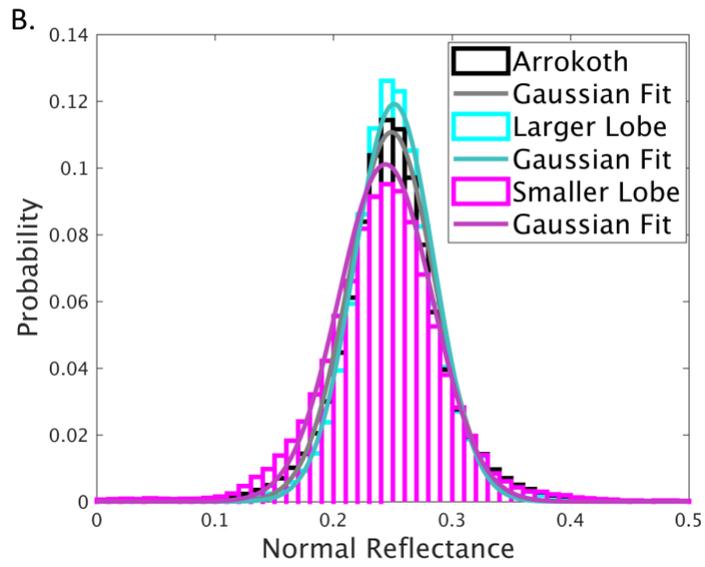
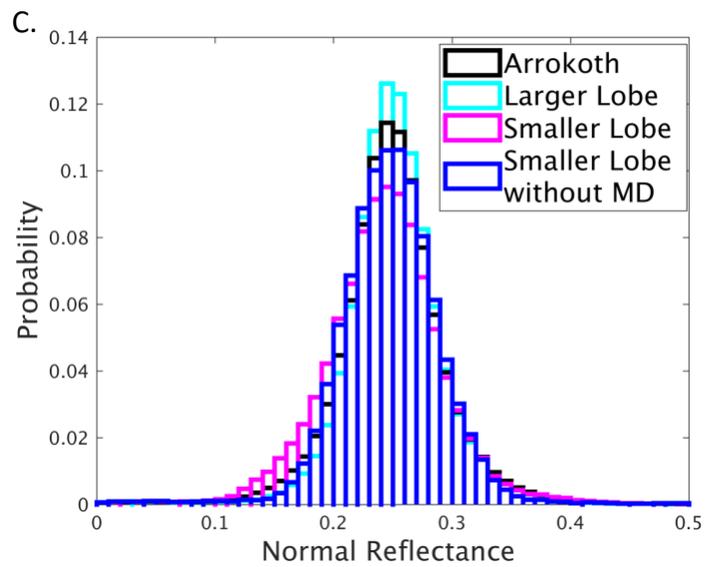

Figure 3: A. Normal reflectance distributions of (486958) Arrokoth and its two lobes at the New Horizons LORRI pivot wavelength of 607.6 nm. The distributions are similar, all three are unimodal with a peak at ≈ 0.25 and approximately symmetric and Gaussian. B. Best-fit Gaussian distributions on top of distributions in A. C. Normal reflectance distribution of the smaller lobe without its large depression (informally called Maryland or MD). The distributions of the two lobes are even more similar when Maryland is neglected.

## 4. Hemispherical Albedo and Bond Albedo of (486958) Arrokoth

Hemispherical albedo is the ratio of the total power scattered by a surface to the incident power (recall that hemispherical albedo is wavelength specific and bolometric hemispherical albedo is the wavelength-integrated ratio) and is crucial for understanding the thermal evolution of the surface. The lunar-Lambert photometric function describes the scattered radiation for all geometries and the total power scattered can be determined by integrating the function over the emission hemisphere. The equation for the hemispherical albedo is:

$$a(i) = \frac{\int_0^{\frac{\pi}{2}} \int_0^{2\pi} I(\theta, \phi) \cos\theta \sin\theta \, d\phi \, d\theta}{\pi F \cos i} \qquad (4)$$

and upon substituting the lunar-Lambert photometric function (equation 1) for $I(\theta, \phi)/F$ and simplifying:

$$a(i) = \frac{A}{\pi} \int_0^{\frac{\pi}{2}} \int_0^{2\pi} \frac{f(\alpha) \cos\theta \sin\theta}{\cos i + \cos\theta} \, d\phi \, d\theta + (1 - A) \qquad (5)$$

where $a(i)$ is the hemispherical albedo, $\theta$ and $\phi$ are the spherical polar angles ($\theta$ is the same as the emission angle, $e$), and the other variables are consistent with the definitions in previous sections ($I(\theta, \phi)$ is scattered intensity, $F$ is solar flux divided by pi, $i$ is incidence angle, $\alpha$ is solar phase angle, $A$ and $f(\alpha)$ are parameters of the lunar-Lambert function). The solar phase angle can be expressed in terms of the other angles using the spherical law of cosines:

$$\cos \alpha = \cos i \cos \theta + \sin i \sin \theta \cos \phi. \tag{6}$$

Note that the above equations indicate that the hemispherical albedo of a general surface depends on the incidence angle of the incident power. This is not a new finding but is an interesting aspect of hemispherical albedo that has been empirically confirmed and is frequently forgotten and/or neglected (Squyres and Veverka, 1981).

The hemispherical albedo of the Lambertian component of the photometric function is analytic and equal to the normal reflectance of the Lambertian component, $1 - A$, independent of incidence angle. To evaluate the integral for the lunar (Lommel-Seeliger) component, a functional form for $f(\alpha)$ is needed. A variety of functions have been proposed for $f(\alpha)$, with varying degrees of complexity and success (e.g., Hapke, 2012). A three-parameter exponential function of the form:

$$f(\alpha) = c_1 e^{c_2 \alpha^{c_3}} \tag{7}$$

gives a good fit to our $f(\alpha)$ values (including $f(0)$) for Arrokoth with $c_1 = 0.46$, $c_2 = -0.57$, and $c_3 = 0.37$. We numerically integrated the lunar component of the hemispherical albedo with this function.

The left panel of figure 4 shows the hemispherical albedo of Arrokoth as a function of incidence angle. The hemispherical albedo at the LORRI pivot wavelength of 607.6 nm monotonically increases from 0.059 at an incidence angle of 0º to a limit 0.086 at 90º (the equation for the hemispherical albedo is not defined at $i = 90$º but the limit exists). This is a nearly 50% increase in the hemispherical albedo over the full range of incidence angles and demonstrates that the variation of hemispherical albedo with incidence angle can be substantial. Note that the hemispherical albedo dependence on incidence angle is not always monotonic, depending on the functional form of $f(\alpha)$. The mean hemispherical albedo of Arrokoth, at 607.6 nm, is 0.063 ± 0.015 and the incidence angle cosine expected value ($a(i) \cos i$) is 0.061 ± 0.015. The uncertainty is an estimate, represents approximately 2-sigma or 68% confidence, and is dominated by the uncertainty of $f(\alpha)$; the formal uncertainty from propagating the uncertainty of the geometric albedo is an order of magnitude smaller. The right panel of figure 4 shows a mean (incidence angle average) hemispherical albedo map of Arrokoth. The map was produced assuming that all locations on the surface of Arrokoth have the same photometric behavior (dependence on incidence, emission, and solar phase angles) but differ in their normal reflectance, consistent with earlier assumptions to determine the best-fit photometric function and normal reflectance map. The hemispherical albedo and normal reflectance maps differ only in their absolute values by the ratio of the mean hemispherical albedo to the mean normal reflectance; the map in figure 4 was produced by multiplying the CA06 map in figure 2 by the ratio of the mean hemispherical albedo to the mean normal reflectance.

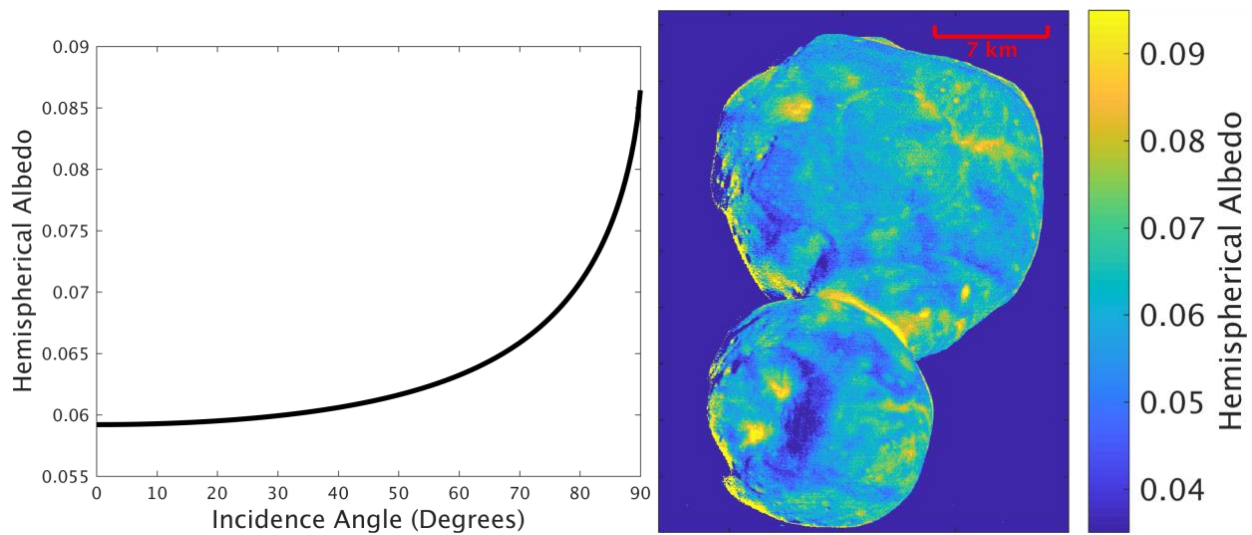

Figure 4: Hemispherical albedo of (486958) Arrokoth at New Horizons LORRI pivot wavelength of 607.6 nm. The left panel shows the hemispherical albedo as a function of incidence angle; the variation with incidence angle is significant. The mean hemispherical albedo is $0.063 \pm 0.015$. The right panel is a map of the mean (incidence angle average) hemispherical albedo. The mean hemispherical albedo map differs from the normal reflectance map in figure 2 only by a multiplicative factor.

Bond albedo is another common photometric parameter in planetary science. It is the ratio of the total power scattered by a planetary body to the incident power; it is different from hemispherical albedo in that it is disk-integrated. Bond albedo can be generalized to a non-spherical shape, but in practice a spherical shape is assumed. Bond albedo does not depend on incidence angle. The Bond albedo of the Lambertian component of the lunar-Lambert photometric function is $1 - A$. We numerically integrated the lunar component of the spherical Bond albedo (Buratti and Veverka, 1983), using the same exponential function for $f(\alpha)$ as above. The Bond albedo for the combined lunar-Lambert photometric function of Arrokoth is $0.062 \pm 0.015$ at 607.6 nm. The Bond albedo of Arrokoth is very similar to its mean hemispherical albedo, but these parameters could differ for some shapes and photometric functions. Note that the visible disk-integrated (assuming a spherical shape) solar phase curve of Arrokoth is presented and analyzed in Stern et al., 2019.

## 5. Comparison with Albedos of Cognate Solar System Objects

Based on its orbital parameters, Arrokoth is likely a cold classical Kuiper belt object (CCKBO; Porter et al., 2018). The orbital parameter phase space of the hot classical KBO (HCKBO) population, however, overlaps with that of the CCKBO population and based on orbital characteristics alone, there is a chance that Arrokoth is a HCKBO (Petit et al., 2011). The albedos of most known KBOs are not strongly constrained, but the geometric albedos of 8 CCKBOs and 25 HCKBOs were determined by fitting a thermal model to radiometric observations (Vilenius et al., 2014; Lacerda et al., 2014). Arrokoth was not included in that sample and the KBOs in that sample were approximately an order of magnitude larger in diameter than Arrokoth. The median V-band geometric albedo of these CCKBOs is 0.15 with a 68% confidence interval of 0.09-0.23 and of the HCKBOs is 0.08 with a 68% confidence interval of 0.04-0.13 (Lacerda et al., 2014). The Arrokoth V-band (pivot wavelength of 551.1 nm) geometric albedo of $\approx 0.21$ is consistent with the CCKBO distribution; it is greater than the median but in the 68% confidence interval. Arrokoth's geometric albedo is less consistent with the HCKBO distribution. Thus, the geometric albedo of Arrokoth adds to the orbital (Porter et al., 2018) and color (Grundy et al., accepted) evidence that Arrokoth is a CCKBO.

Aside from Arrokoth, the only KBOs to be explored by a spacecraft are Pluto and its satellites. Pluto's geometric albedo varies significantly with sub-observer latitude and longitude due to extreme variations of the normal reflectance of its surface; the mean geometric albedo at the LORRI pivot wavelength is 0.62 (Buratti et al., 2017). Charon's geometric albedo at the LORRI pivot wavelength is 0.41 (Buratti et al., 2017) whereas Arrokoth's is ≈ 0.24. Pluto and Charon have much greater geometric albedos than Arrokoth, as well as other KBOs in general. Pluto and other large KBOs such as Eris and Makemake have characteristically high geometric albedos due to the retention of volatiles on their surface (e.g., Schaller and Brown, 2007). Charon has not retained its surface volatiles, but it is also larger than most KBOs and its surface is geologically evolved (Moore et al., 2016) so it's unsurprising that the geometric albedos of both Pluto and Charon differ from that of Arrokoth. The Pluto system's smaller satellites have geometric albedos from 0.56-0.83 (Weaver et al., 2016). These satellites are more similar in size to Arrokoth, but likely formed as the result of a giant impact (e.g., McKinnon et al., 2017), which resulted in a different composition and history than Arrokoth and most other KBOs. Thus, Arrokoth's geometric albedo is different from all of the bodies in the Pluto system.

Arrokoth's normal reflectance, however, is similar to specific dark regions on Pluto and Charon. Pluto has at least two distinct dark terrains: (1) the darkest (and reddest) terrain at the equator including Cthulhu (informal name) and (2) a terrain that is also relatively dark but distinct from the equatorial dark regions in its normal reflectance distribution (it is also less red) in Viking Terra (Buratti et al., 2017; Olkin et al., 2017). The normal reflectance of the latter is similar to Arrokoth. Mordor Macula (informal name) at Charon's north pole is its darkest observed terrain and has a mean normal reflectance similar to Arrokoth. The dark, red materials of Viking Terra and Mordor Macula are thought to be organic macromolecules produced by energetic radiation processing of hydrocarbons (tholins; Grundy et al., 2016a; 2016b). The similar normal reflectance of these surfaces could be a result of similar initial compositions and radiolytic evolution within the Kuiper belt.

Saturn's satellite Phoebe and other irregular satellites of the giant planets are hypothesized to be captured and possibly former KBOs (e.g., Johnson and Lunine, 2005). The V-band geometric albedo of Phoebe is 0.09 (Miller et al., 2011) and the irregular satellites generally have geometric albedos ≤ 0.10 (e.g., Grav et al., 2015; Triton is an exception because it is large enough to retain its volatiles). Centaurs and Jupiter family comets are also hypothesized to be former KBOs that have been scattered out of the Kuiper belt. Centaurs are bimodal in color and albedo (e.g., Bauer et al., 2013): the darker/grayer and brighter/redder groups have mean visible geometric albedos

(determined by fitting thermal models to radiometric observations) of 0.06 ± 0.02 and 0.12 ± 0.05, where the uncertainties indicate one standard deviation. One Centaur, (145486) 2005 UJ438, has a modeled visible geometric albedo of > 0.20 in both WISE/NEOWISE-based (Bauer et al., 2013) and Herschel-based (Duffard et al., 2014) analyses. But, both analyses also report large uncertainties for this Centaur. Two other Centaurs have modeled visible geometric albedos > 0.18 in WISE/NEOWISE-based analysis (Bauer et al., 2013), but substantially lower values in Herschel-based analysis (Duffard et al., 2014). The median modeled geometric albedo of a sample of 24 Jupiter family comets is 0.042 with a standard deviation of 0.013 (Kokotanekova et al., 2017). 67P/Churyumov–Gerasimenko, the best explored Jupiter family comet, has a geometric albedo of 0.062 at 550 nm (Ciarniello et al., 2015). Thus, the irregular satellites, darker and grayer group Centaurs, and Jupiter family comets are significantly darker than CCKBOs and this is also true for the best explored objects in each family: Phoebe, 67P/Churyumov–Gerasimenko, and Arrokoth (Centaurs have not yet been explored with spacecraft). This result suggests that they originate from a different population and/or were darkened after their departure from the Kuiper belt.

## 6. Conclusions

Kuiper belt object (486958) Arrokoth has a geometric albedo of $p_v = 0.21^{+0.05}_{-0.04}$ at a wavelength of 550 nm and ≈ 0.24 at 610 nm. Its geometric albedo is greater than the median of, but consistent with, a distribution of cold classical KBOs, and is less consistent with the hot classical KBO distribution, which adds to the orbital (Porter et al., 2018) and color (Grundy et al., accepted) evidence that Arrokoth is a cold classical KBO. Thus, Arrokoth may be the most primitive object explored by a spacecraft.

The normal reflectance of Arrokoth's surface varies with location, ranging from ≈ 0.10 – 0.40 at a pivot wavelength of 610 nm with an approximately Gaussian distribution. The normal reflectance distributions of Arrokoth's two lobes are similar, both are approximately Gaussian, peak at ≈ 0.25 at 610 nm, and range from ≈ 0.10 – 0.40. The photometric similarity of the two lobes is consistent with co-formation and co-evolution.

The hemispherical albedo of Arrokoth varies substantially with both incidence angle and location, the average is 0.063 ± 0.015 at 610 nm. The Bond albedo of Arrokoth is 0.062 ± 0.015 at 610 nm.

Acknowledgements: We are sincerely grateful to the entire New Horizons team for enabling this research. We thank the NASA New Horizons project for financial support. J.D.H. gratefully acknowledges financial support from the NASA Postdoctoral Program. This research was carried out at the Jet Propulsion Laboratory, California Institute of Technology, under a contract with the National Aeronautics and Space Administration. We thank Stefano Mottola, an anonymous referee, and editor Will Grundy for their service and helpful comments.